\documentclass[aps,pra,amsmath,showpacs,twocolumn,floatfix,superscriptaddress]{revtex4}
\usepackage{bm}
\usepackage{graphicx}
\begin{document}
\title{Enhancing quantum entanglement for continuous variables by a coherent superposition of photon subtraction and addition}

\author{Su-Yong Lee}
\affiliation{Department of Physics, Texas A\&M University at Qatar,
 POBox 23874, Doha, Qatar}

\author{Se-Wan Ji}
\affiliation{School of Computational Sciences, Korea Institute for Advanced Study, Seoul 130-012, South Korea}

\author{Ho-Joon Kim}
\affiliation{Department of Physics, Texas A\&M University at Qatar,
 POBox 23874, Doha, Qatar}

\author{Hyunchul Nha}
\affiliation{Department of Physics, Texas A\&M University at Qatar,
 POBox 23874, Doha, Qatar}
\affiliation{Institute f{\"u}r Quantenphysik, Universit{\"a}t Ulm, D-89069 Ulm, Germany}

\begin{abstract}
We investigate how the entanglement properties of a two-mode state can be improved by performing a coherent superposition operation $t\hat{a}+r\hat{a}^\dag$ of photon subtraction and addition, proposed by Lee and Nha [\pra {\bf 82}, 053812 (2010)], on each mode.
We show that the degree of entanglement, the EPR-type correlation, and the performance of quantum teleportation can be all enhanced for the output state when the coherent operation is applied to a two-mode squeezed state. The effects of the coherent operation are more prominent than those of the mere photon subtraction $\hat{a}$ and the addition $\hat{a}^\dag$ particularly in the small squeezing regime, whereas the optimal operation becomes the photon subtraction (case of $r=0$) in the large-squeezing regime.
\end{abstract}

\pacs{03.67.-a, 42.50.Dv}
\maketitle

\section{Introduction}
Continuous variable (CV) entangled resources are essential for many applications in quantum information processing, such as long distance quantum communications \cite{Briegel}, CV quantum teleportation \cite{Vaidman, Braunstein}, and the Bell test \cite{Nha, Garcia,Reid,Sewan}.
As the practically available resources usually have finite degree of entanglement, it is of crucial importance to come up with feasible, mostly probabilistic, schemes to improve the given entanglement properties. In the regime of CVs, a Gaussian two-mode entangled state is most frequently employed as an entangled resource both theoretically and experimentally.
It is known that the Gaussian entangled states cannot be distilled only by Gaussian local operations and classical communications
\cite{Eisert, Fiurasek, Giedke}, but they can be made so by non-Gaussian operations.
A simple non-Gaussian operation is the photon subtraction, represented by $\hat{a}|\Psi\rangle$ where $\hat{a}$ is the bosonic annihilation operator, which can enhance the performance of the Gaussian two-mode entangled state in quantum teleportation \cite{Opatrny, Cochrane, Olivares} and the dense-coding \cite{Kitagawa}.
 Entanglement improvement was experimentally realized by a nonlocal photon subtraction \cite{Alexei} and by the local photon subtractions \cite{Takahashi}, respectively. Furthermore, it was suggested that entanglement distillation can be achieved also by photon addition, represented by $\hat{a}^\dag|\Psi\rangle$ where $\hat{a}^\dag$ is the bosonic creation operator \cite{Yang}, or  by some combinations of addition and subtraction, such as photon-addition-then-subtraction $\hat{a}\hat{a}^\dag|\Psi\rangle$ and subtraction-then-addition $\hat{a}^\dag \hat{a}|\Psi\rangle$ \cite{Yang}, or by coherent combinations of two sequences of photon subtraction and addition \cite{Fiurasek1}.
 In particular, the latter scheme was motivated by the recent proposal of M. S. Kim  {\it et al.} to prove the bosonic commutation relation  \cite{Kim1}, where the coherent superpositions of two product operations, $\hat{a}\hat{a}^\dag$ and $\hat{a}^\dag \hat{a}$, was suggested, which was experimentally demonstrated in \cite{Zavatta1}.

Recently, we proposed a coherent superposition of photon subtraction and addition at a more elementary level than Ref. \cite{Kim1}, that is, $t\hat{a}+r\hat{a}^\dag$ for quantum state engineering \cite{Lee}. The coherent operation can be realized in a single photon interference setting \cite{Kim1,Lee}, where the key idea is to erase the which-path information on whether the single photon triggering the detector is due to the experimental configuration implementing the subtraction or the addition.
The coherent operation may create a nonclassical state with observable effects (squeezing and sub-Poissonian statistics) out of a classical state and can also be used to generate an arbitrary superposition states in the number-state basis, e.g. $C_0|0\rangle+C_1|1\rangle+C_2|2\rangle$ \cite{Lee}.

In this paper, we apply the coherent superposition operation $t\hat{a}+r\hat{a}^\dag$ to two-mode states and study how the entanglement properties can be enhanced by the coherent operation on each local mode. This naturally includes the photon subtraction ($r=0$) and the photon addition ($t=0$) as extremal cases. Specifically, we investigate how (i) the degree of entanglement quantified by von Neuman entropy for pure states, (ii) the Einstein-Podolsky-Rosen (EPR) correlation, and (iii) the performance (average fidelity) for quantum teleportation can be improved by acting the coherent operation on a two-mode squeezed state. We show that all these entanglement characteristics can be remarkably enhanced by the coherent operation particularly in the small-squeezing regime, whereas the optimal operation becomes the photon subtraction (case of $r=0$) in the large-squeezing regime.

This paper is organized as follows. In Sec. II, we study how the coherent superposition operation transforms a two-mode squeezed state to a non-Gaussian state with the enhancement of entanglement measured by von Neumann entropy and that of EPR correlation measured by the sum of two nonlocal quadrature variances. In Sec. III, we employ the output non-Gaussian entangled state via the coherent operation for CV teleportation, where the average fidelity is compared with those obtained by other non-Gaussian operations on two-mode states. The main results are summarized in Sec. IV.

\section{Non-Gaussian two-mode entangled states by a coherent superposition of photon subtraction and addition}
In this section, we study how the local coherent operation $t\hat{a}+r\hat{a}^\dag$ on two-mode states can change the entanglement properties, specifically the degree of entanglement and the EPR correlation, of the input states. In particular, we take a two-mode squeezed state as an input state, which is a typical Gaussian state produced in experiment. It is also known that every Gaussian pure state can be transformed to a two-mode squeezed state by local unitary Gaussian operations \cite{Laurat}. It has been previously found that the entanglement can be distilled by some non-Gaussian operations such as photon subtraction (addition) and addition-then-subtraction (subtraction-then-addition) operations. We show that the entanglement can be even more enhanced by a coherent superposition operation particularly in the low-squeezing regime.

By performing the coherent operation on the two-mode squeezed state, $|TMSS\rangle_{AB}= \sqrt{1-\lambda^2}\sum^{\infty}_{n=0}\lambda^n|n\rangle_A|n\rangle_B$ ($\lambda=\tanh{s}$), one obtains the output state as
\begin{widetext}
\begin{eqnarray}
&&\sqrt{N}(t_A\hat{a}+r_A\hat{a}^{\dag})(t_B\hat{b}+r_B\hat{b}^{\dag})|TMSS\rangle_{AB}, \nonumber\\
&&=\sqrt{N}\left[\right.\sum^{\infty}_{n=0}\lambda^n (\lambda t_At_B (n+1)|n\rangle_A|n\rangle_B
+\lambda t_Ar_B\sqrt{(n+1)(n+2)}|n\rangle_A|n+2\rangle_B+\lambda r_At_B\sqrt{(n+1)(n+2)}|n+2\rangle_A|n\rangle_B\nonumber\\
&&+r_Ar_B (n+1)|n+1\rangle_A|n+1\rangle_B)\left.\right],
\end{eqnarray}
\end{widetext}
where
\begin{eqnarray}
N=\frac{(1-\lambda^2)^3}{\lambda^2(1+|t_Ar^\ast_B+r_At^\ast_B|^2)+|t_At_B\lambda^2+r_Ar_B|^2}
\end{eqnarray}
 is the normalization constant with $|t_i|^2+|r_i|^2=1$ ($i=A,B$).
We examine the entanglement and the EPR correlation of the output state compared with those states by other non-Gaussian operations: photon subtraction operation $\hat{a}_i$ on one (both) mode(s) of the TMSS, photon-addition-then-subtraction operation $\hat{a}_i\hat{a}_i^\dag$ on both modes of the TMSS. For the case of photon addition $\hat{a}^{\dag}\hat{b}^{\dag}|TMSS\rangle$, the degree of entanglement is the same as that of the photon-subtracted state $\hat{a}\hat{b}|TMSS\rangle$ , but it does not improve the EPR correlation \cite{Yang}. Thus, we do not here consider the states $\hat{a}^{\dag}\hat{b}^{\dag}|TMSS\rangle$. 
Those non-Gaussian operations on the TMSS generate the output states as
\begin{eqnarray}
\hat{a}|TMSS\rangle_{AB}&\dot{=}&\hat{b}^{\dag}|TMSS\rangle_{AB}\nonumber\\
&\dot{=}&\sqrt{m_1}\sum^{\infty}_{n=0}\lambda^n\sqrt{n+1}|n\rangle_A|n+1\rangle_B,\nonumber\\
\hat{b}|TMSS\rangle_{AB}&\dot{=}&\hat{a}^{\dag}|TMSS\rangle_{AB}\nonumber\\&\dot{=}&
\sqrt{m_1}\sum^{\infty}_{n=0}\lambda^n\sqrt{n+1}|n+1\rangle_A|n\rangle_B,\nonumber\\
\hat{a}\hat{b}|TMSS\rangle_{AB}&\dot{=}&\hat{a}\hat{a}^{\dag}|TMSS\rangle_{AB}\nonumber\\
&\dot{=}&\sqrt{m_2}\sum^{\infty}_{n=0}\lambda^n(n+1)|n\rangle_A|n\rangle_B, \nonumber\\
\hat{a}\hat{a}^{\dag}\hat{b}\hat{b}^{\dag}|TMSS\rangle_{AB}&&\nonumber\\
&\dot{=}&\sqrt{m_3}\sum^{\infty}_{n=0}\lambda^n (n+1)^2|n\rangle_A|n\rangle_B,
\end{eqnarray}
where $\dot{=}$ represents states equal up to the normalization constants $m_1=(1-\lambda^2)^2$, $m_2=\frac{(1-\lambda^2)^3}{1+\lambda^2}$, and
$m_3=\frac{(1-\lambda^2)^5}{1+11\lambda^2+11\lambda^4+\lambda^6}$.

\subsection{Degree of Entanglement}
We here investigate how the degree of entanglement is changed by the local coherent operations $t\hat{a}+r\hat{a}^\dag$ on a two-mode state.
For a pure state in Schmidt form, $|\psi\rangle_{AB}=\sum_{i=1}c_i|\alpha_i\rangle_A|\beta_i\rangle_B$ ($c_i$: real positive) with the orthonormal states $|\alpha_i\rangle_A$ and $|\beta_i\rangle_B$, the quantum entanglement is quantified by the entropy of the reduced density operator \cite{Bennett},
\begin{eqnarray}
E(|\psi\rangle_{AB})=-{\rm Tr}\rho_{A}\log_2\rho_{A}=-\sum_i c^2_i\log_2 c^2_i,
\end{eqnarray}
where the local state is given by $\rho_A=Tr_B|\psi\rangle_{AB}\langle\psi|$.
The amount of entanglement for a two-mode squeezed state is analytically given by $E=\cosh^2{s}\log_2(\cosh^2{s})-\sinh^2{s}\log_2(\sinh^2{s})$ \cite{Enk,Ryu}, and those for other states can be evaluated numerically by their Schmidt coefficients.

In Fig. 1, we plot the degree of entanglement for the four states $|TMSS\rangle$,
$\hat{a}|TMSS\rangle$, $\hat{a}\hat{b}|TMSS\rangle$, and $\hat{a}\hat{a}^{\dag}\hat{b}\hat{b}^{\dag}|TMSS\rangle$.  It is seen that the entanglement is best improved by the photon-addition-then-subtraction operation $\hat{a}\hat{a}^{\dag}\hat{b}\hat{b}^{\dag}$ among those. However, the entanglement is even more improved by the coherent superposition operation $t\hat{a}+r\hat{a}^\dag$ particularly in the weak squeezing region with the local parameters the same $r_A=r_B=r$, as shown in Fig. 1 (a) and (b).  This may be understood by looking into their Schmidt coefficients. Assuming $\lambda=\tanh{s}\ll1$, the (unnormalized) states are given by
\begin{eqnarray}
&&\hat{a}\hat{b}\hat{a}^{\dag}\hat{b}^{\dag}|TMSS\rangle \approx  |0\rangle_A|0\rangle_B+4\lambda|1\rangle_A|1\rangle_B, \nonumber\\\nonumber\\
&&(t\hat{a}+r\hat{a}^{\dag})|TMSS\rangle \nonumber\\
&&\approx  r|1\rangle_A|0\rangle_B+\lambda (t|0\rangle_A+ \sqrt{2}r|2\rangle_A)|1\rangle_B,\nonumber\\\nonumber\\
&&(t\hat{a}+r\hat{a}^{\dag})(t\hat{b}+r\hat{b}^{\dag})|TMSS\rangle \nonumber\\
&&\approx  r^2|1\rangle_A|1\rangle_B+\lambda (t|0\rangle_A+\sqrt{2}r|2\rangle_A)(t|0\rangle_B+\sqrt{2}r|2\rangle_B).\nonumber\\
\end{eqnarray}
We then obtain the corresponding Schmidt coefficients as shown in the table below.

\begin{tabular}{|c|c|}
\hline state & Schmidt coefficient \\
\hline \tiny{$\hat{a}\hat{b}\hat{a}^{\dag}\hat{b}^{\dag}|TMSS\rangle $}  & $\frac{1}{\sqrt{1+16\lambda^2}}$, $\frac{4\lambda}{\sqrt{1+16\lambda^2}}$\\
\hline \tiny{$(t\hat{a}+r\hat{a}^{\dag})|TMSS\rangle$} & $\frac{r}{\sqrt{r^2+\lambda^2(1+r^2)}}$, $\frac{\lambda\sqrt{1+r^2}}{\sqrt{r^2+\lambda^2(1+r^2)}}$\\
\hline \tiny{$(t\hat{a}+r\hat{a}^{\dag})(t\hat{b}+r\hat{b}^{\dag})|TMSS\rangle$}& $\frac{r^2}{\sqrt{r^4+\lambda^2(1+r^2)^2}}$,
$\frac{\lambda(1+r^2)}{\sqrt{r^4+\lambda^2(1+r^2)^2}}$\\
\hline
\end{tabular}

In the limit of $\lambda\ll1$, one Schmidt coefficient of the state $\hat{a}\hat{b}\hat{a}^{\dag}\hat{b}^{\dag}|TMSS\rangle$ is always much larger than the other coefficient. On the other hand, it is possible to equalize the Schmidt coefficients for the case of the coherent superposition operation, which explains  the larger improvement of the output entanglement than the case of $\hat{a}\hat{b}\hat{a}^{\dag}\hat{b}^{\dag}$.
In Fig. 1(b), the entanglement achieves around $1$ for $s=0.1$ via the coherent superposition operation on one mode or both modes of the TMSS, whereas it achieves around 0.6 via the photon-addition-then-subtraction operation. At $r=0$ ($r=1$), the state refers to the photon-subtracted state (the photon-added state).
\begin{figure}
\centerline{\scalebox{0.35}{\includegraphics[angle=90]{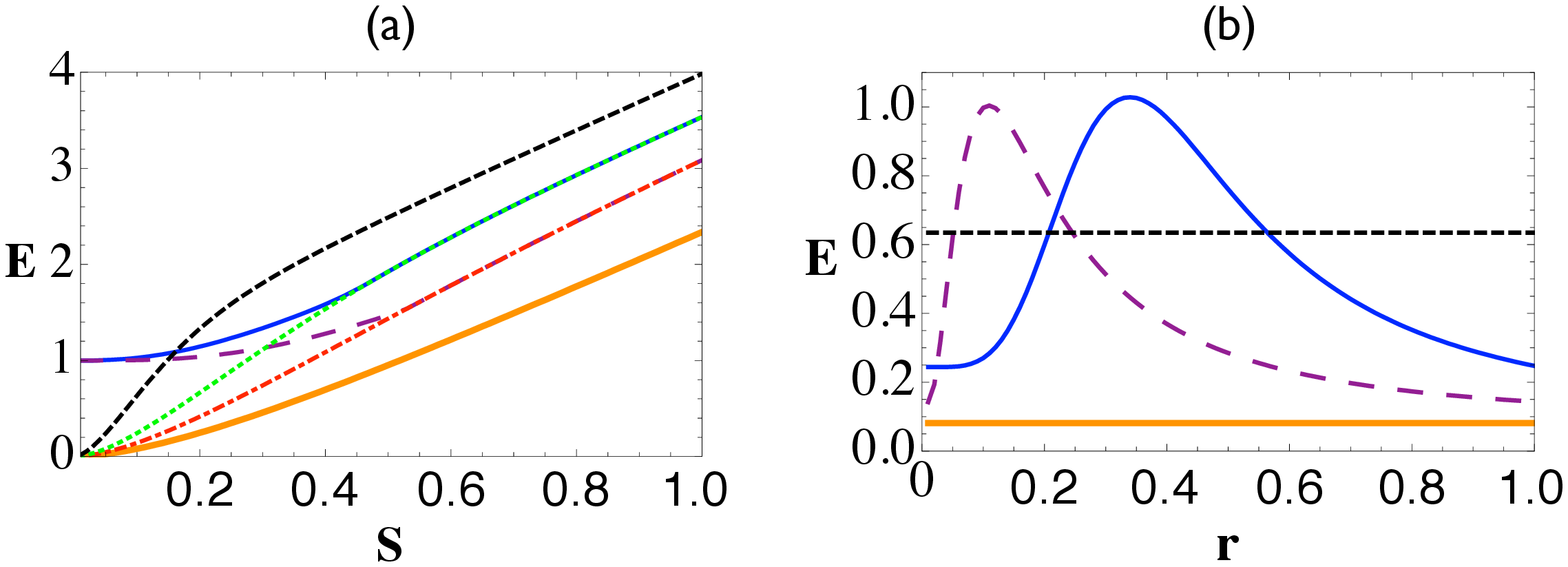}}}
\vspace{-1.5in}
\caption{(Color online) Entanglement (a) as a function of the squeezing parameter $s$, and (b) as a function of $r$ with $s=0.1$ for the states: $|TMSS\rangle$ (orange thick solid) , $\hat{a}|TMSS\rangle$ (red dot-dashed), $\hat{a}\hat{b}|TMSS\rangle$ (green dotted), $\hat{a}\hat{a}^{\dag}\hat{b}\hat{b}^{\dag}|TMSS\rangle$ (black dashed), $(t\hat{a}+r\hat{a}^{\dag})|TMSS\rangle$ (the coherent operation on one local mode only, purple long-dashed), and $(t\hat{a}+r\hat{a}^{\dag})(t\hat{b}+r\hat{b}^{\dag})|TMSS\rangle$ (the coherent operation on both modes, blue solid). In (a), the value $r$ in the coherent operation is optimized for each $s$. }
\label{fig:fig1}
\end{figure}

\subsection{EPR correlation}
We here investigate another entanglement property, the EPR correlation, which is the total variance of a pair of EPR-like operators, $\Delta^2 (\hat{x}_A-\hat{x}_B)+\Delta^2 (\hat{p}_A+\hat{p}_B)$, where $\hat{x}_j=\frac{1}{\sqrt{2}}(\hat{a}_j+\hat{a}^{\dag}_j)$ and $\hat{p}_j=\frac{1}{i\sqrt{2}}(\hat{a}_j-\hat{a}^{\dag}_j)$ ($j=A,B$). For separable two-mode states, the total variance is larger than or equal to $2$, so
the condition $\Delta^2 (\hat{x}_A-\hat{x}_B)+\Delta^2 (\hat{p}_A+\hat{p}_B)<2$ clearly indicates quantum entanglement \cite{Duan}, which can be a crucial resource for quantum protocols using CVs.

On applying the coherent operation,  $(t_A\hat{a}+r_A\hat{a}^{\dag})(t_B\hat{b}+r_B\hat{b}^{\dag})|TMSS\rangle$, we obtain the EPR correlation
\begin{eqnarray}
&&\Delta^2 (\hat{x}_A-\hat{x}_B)+\Delta^2 (\hat{p}_A+\hat{p}_B)\nonumber\\
&&=2+\frac{4}{M}[ M(\cosh{s}-\sinh{s})(\cosh{s}-2\sinh{s})\nonumber\\
&&-(AB+|B|^2)(\cosh{s}-\sinh{s})^2],
\end{eqnarray}
where
\begin{eqnarray}
A&=&t_At_B\sinh^2{s}+r_Ar_B\cosh^2{s},\nonumber\\
B&=&(t_At_B+r_Ar_B)\cosh{s}\sinh{s},\nonumber\\
C&=&\sqrt{2}t_Ar_B\cosh{s}\sinh{s},\nonumber\\
D&=&\sqrt{2}r_At_B\cosh{s}\sinh{s},
\end{eqnarray}
and $M=A^2+B^2+C^2+D^2$.
In Fig. 2, we plot the EPR correlation for the output state and see that there exists a threshold curve as a function of $s$ and $r$ ($r_A=r_B=r$).
At $r=0$ ($r=1$), the case refers to the photon-subtracted (photon-added) state, $\hat{a}\hat{b}|TMSS\rangle$ ($\hat{a}^{\dag}\hat{b}^{\dag}|TMSS\rangle$). To exhibit the EPR correlation, the squeezing parameter must be larger than $s=0~(0.3782)$ at $r=0~(1)$.
The EPR correlation is stronger in the photon-subtracted state than in the photon-added state for the same squeezing parameter.
However, for a given $s$, the optimal EPR correlation occurs with the coherent superposition operation, $r\neq 0,1$, in the weak squeezing region, and the optimized $r$ approaches to $r=0$ (photon subtraction) with the squeezing parameter $s$.

\begin{figure}
\centerline{\scalebox{0.25}{\includegraphics[angle=90]{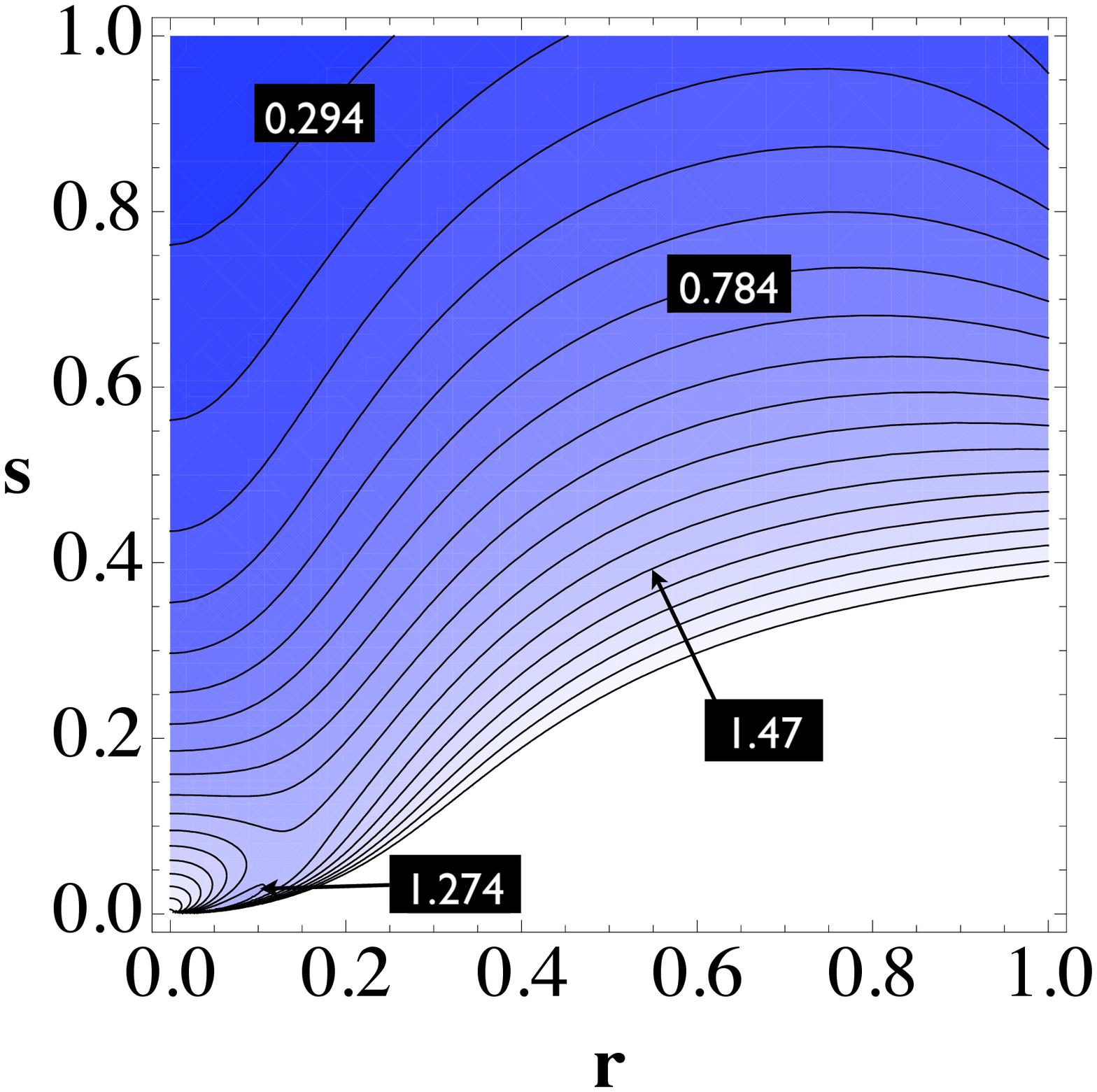}}}
\vspace{-0.2in}
\caption{(Color online) EPR correlation $\Delta^2 (\hat{x}_A-\hat{x}_B)+\Delta^2 (\hat{p}_A+\hat{p}_B)$ as a function of $s$ and $r$ for the state $(t\hat{a}+r\hat{a}^{\dag})(t\hat{b}+r\hat{b}^{\dag})|TMSS\rangle$. Colored region represents the condition $\Delta^2 (\hat{x}_A-\hat{x}_B)+\Delta^2 (\hat{p}_A+\hat{p}_B)<2$.}
\label{fig:fig2}
\end{figure}

In Fig. 3 (a), we show the EPR correlation from the coherent superposition operation optimized over the ratio $r$ in $t\hat{a}+r\hat{a}^{\dag}$  for a given squeezing parameter s, compared with those from the other non-Gaussian operations. We see that the coherent operation on both local modes improves the EPR correlation better than the other operations in the weak squeezing region;
For the class of photon-number entangled states $\sum_{n=0}^{N}d_n|n\rangle_A|n\rangle_B$ with the truncation number $N$, the EPR correlation is given by
\begin{eqnarray}
&&\Delta^2 (\hat{x}_A-\hat{x}_B)+\Delta^2 (\hat{p}_A+\hat{p}_B) \nonumber\\
&&=2-4\frac{\sum^N_{n=1}n(d_{n-1}-d_n)d_n}{\sum^N_{n=0}d_n^2},
\end{eqnarray}
where $d_n$ is taken as real. For a given truncation $N$, the EPR correlation optimized over the coefficients $\{d_n\}$ becomes stronger with $N$, as shown in Fig. 4.
To the first order of $\lambda$, the effective $N$ becomes larger by the coherent operation than by the other non-Gaussian operations.
From Eq.~(5), $N_{\rm eff}$ reads 1 and 2 for the states $\hat{a}\hat{b}\hat{a}^{\dag}\hat{b}^{\dag}|TMSS\rangle$ and $(t\hat{a}+r\hat{a}^{\dag})(t\hat{b}+r\hat{b}^{\dag})|TMSS\rangle$, respectively. Although the state $(t\hat{a}+r\hat{a}^{\dag})(t\hat{b}+r\hat{b}^{\dag})|TMSS\rangle$ have other component states $|20\rangle$ and $|02\rangle$ in Eq.~(5), we have checked that the optimized EPR correlation including these states is the same as that of Eq.~(9) with $N=2$ .
For the states $d_0|00\rangle_{AB}+d_1|11\rangle_{AB}+d_2|22\rangle_{AB}$, the optimized condition reads $d_0\approx4.51,~d_1\approx2.63$, and $d_2\approx1.15$. The coherent operation does not exactly achieve the optimal EPR correlation, but one can adjust the parameter $r$ in the coherent operation $t\hat{a}+r\hat{a}^{\dag}$ to outperform the other non-Gaussian operations.

\begin{figure}
\centerline{\scalebox{0.35}{\includegraphics[angle=90]{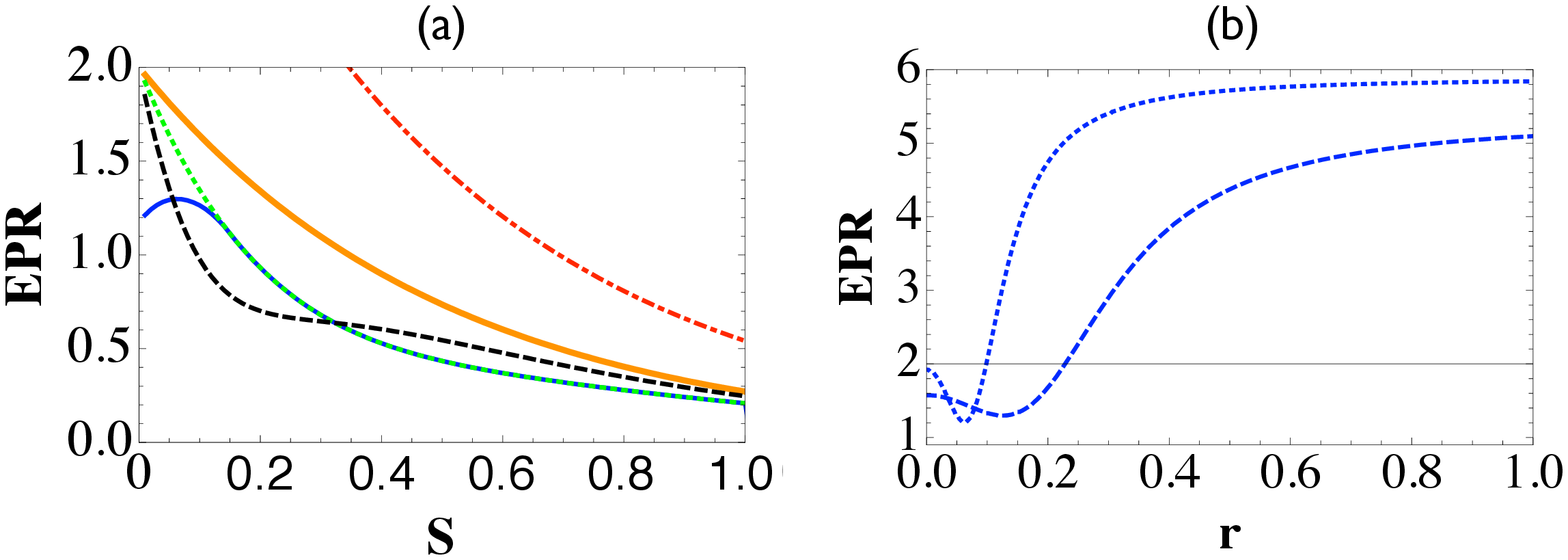}}}
\vspace{-1.5in}
\caption{(Color online) (a) EPR correlation as a function of $s$ for the states:$|TMSS\rangle$ (orange thick solid), $\hat{a}|TMSS\rangle$ (red dot-dashed), $\hat{a}\hat{b}|TMSS\rangle$ (green dotted), $\hat{a}\hat{a}^{\dag}\hat{b}\hat{b}^{\dag}|TMSS\rangle$ (black dashed), and $(t\hat{a}+r\hat{a}^{\dag})(t\hat{b}+r\hat{b}^{\dag})|TMSS\rangle$ (blue solid),
(b) EPR correlation as a function of $r$ for the state, $(t\hat{a}+r\hat{a}^{\dag})(t\hat{b}+r\hat{b}^{\dag})|TMSS\rangle$ at $s=0.01$ (blue dotted) and $s=0.06$ (blue dashed). In (a), the value $r$ in the coherent operation is optimized for each $s$.}
\label{fig:fig3}
\end{figure}

In general, the EPR correlation of the TMSS is enhanced with the squeezing parameter $s$, but it may not be always true for the case of the coherent operation in the weak squeezing region, as shown in Fig. 3 (a). We particularly compare the cases of $s=0.01$ and $s=0.06$ in Fig. 3 (b).
For the state of the form $\sum_{n=0}^{2}d_n|n\rangle_A|n\rangle_B$, the best EPR correlation is obtained with the condition $d_0\approx4.51,~d_1\approx2.63$, and $d_2\approx1.15$  where the coefficients decrease with the photon number.
For the case of coherent operation $t\hat{a}_i+r\hat{a}_i^{\dag}$ on the TMSS, the coefficients in the number state basis (with the parameter $r$ adjusted for a best result) are more deviated from the optimal condition with increasing $s$. For a moderate squeezing $0.055<s<0.324$, the operation $\hat{a}\hat{a}^{\dag}\hat{b}\hat{b}^{\dag}$ gives the best EPR correlation, whereas for a rather large squeezing $s>0.324$ the photon subtraction ($r=0$) becomes the optimal operation.

Among the considered non-Gaussian operations, the photon subtraction on one mode only does not distill the EPR correlation as shown in Fig. 3 (a), which can also be explained in a similar way. In Fig. 4, the optimal EPR correlation for the class of states $\sum^N_{n=0}e_n|n\rangle_A|n+1\rangle_B$ (like the state $\hat{a}|TMSS\rangle_{AB}$) cannot be better than that for the states $\sum^N_{n=0}d_n|n\rangle_A|n\rangle_B$ for a given $N$.
Note that the non-Gaussian operations $\hat{a}^{\dag}\hat{b}^{\dag}$ and $\hat{a}^{\dag}\hat{a}\hat{b}^{\dag}\hat{b}$ do not enhance the EPR correlation of the TMSS, either \cite{Yang}.

\begin{figure}
\centerline{\scalebox{0.5}{\includegraphics{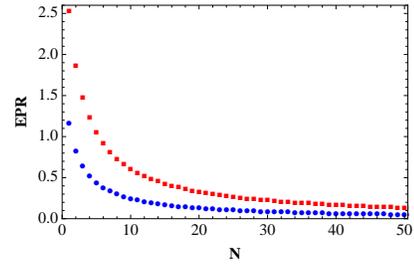}}}
\vspace{-0.1in}
\caption{(Color online) EPR correlation as a function of the truncation number $N$ for the states $\sum_{n=0}^{N}d_n|n\rangle_A|n\rangle_B$ (blue circle) and $\sum_{n=0}^{N}e_n|n\rangle_A|n+1\rangle_B$ (red square) optimized over $\{d_n\}$ and $\{e_n\}$, respectively.}
\label{fig:fig4}
\end{figure}

\section{Quantum teleportation using non-Gaussian entangled states}
In this section, we investigate how the non-Gaussian entangled state via the coherent operation can improve the average fidelity for CV quantum teleportation.
After L. Vaidman introduced the quantum teleportation for the state of a one-dimensional particle in phase space \cite{Vaidman}, it has been further extended to an experimentally relevant protocol using the finite degrees of correlation (squeezing) by Braunstein and Kimble (BK) \cite{Braunstein}. The quality of the teleportation is usually assessed by the average fidelity between an unknown input state and its teleported output state.
The best possible fidelity in teleporting a coherent state without entangled resources is 1/2 \cite{Braunstein2}, so the fidelity over the classical bound 1/2 may be considered as a success for CV quantum teleportation. Based on the BK protocol, the perfect teleportation can occur with an infinitely entangled resource that exhibits an ideal EPR correlation, i.e., $\Delta(\hat{x}_A-\hat{x}_B)\rightarrow 0,~\Delta(\hat{p}_A+\hat{p}_B)\rightarrow 0$.
The CV teleportation is practically implemented with the quadrature amplitudes $\hat{x}=\frac{1}{\sqrt{2}}(\hat{a}+\hat{a}^{\dag})$ and $\hat{p}=\frac{1}{i\sqrt{2}}(\hat{a}-\hat{a}^{\dag})$ of the optical field playing the roles of position and momentum \cite{Braunstein}.

Some non-Gaussian entangled states were previously employed as a quantum resource for CV teleportation \cite{Yang, DellAnno}. We here consider the state  $(t_A\hat{a}+r_A\hat{a}^{\dag})(t_B\hat{b}+r_B\hat{b}^{\dag})|TMSS\rangle$ via the coherent operation for the teleportation of coherent states.
The characteristic function of the state $(t_A\hat{a}+r_A\hat{a}^{\dag})(t_B\hat{b}+r_B\hat{b}^{\dag})|TMSS\rangle$ is given by
\begin{widetext}
\begin{eqnarray}
C_E(\lambda_2,\lambda_3)&=&\frac{e^{-(|\alpha|^2+|\beta|^2)/2}}{M}[A^2(1-|\alpha|^2)(1-|\beta|^2)+B^2
+C^2(1-2|\beta|^2+\frac{|\beta|^4}{2})+D^2(1-2|\alpha|^2+\frac{|\alpha|^4}{2})\nonumber\\
&+&AB(\alpha\beta+\alpha^*\beta^*)+\frac{A}{\sqrt{2}}(\alpha\beta^*+\alpha^*\beta)
\{C(|\beta|^2-2)+D(|\alpha|^2-2)\}+\frac{BC}{\sqrt{2}}(\beta^2+\beta^{*2})\nonumber\\
&+&\frac{BD}{\sqrt{2}}(\alpha^2+\alpha^{*2})
+\frac{CD}{2}(\alpha^2\beta^{*2}+\alpha^{*2}\beta^2)],
\end{eqnarray}
\end{widetext}
where $\alpha=\lambda_2\cosh{s}-\lambda^*_3\sinh{s}$, and $\beta=\lambda_3\cosh{s}-\lambda^*_2\sinh{s}$. In general, with the characteristic function $C_E(\lambda_2,\lambda_3)$ for the entangled resource, the characteristic function $C_{\rm out}(\lambda)$ of the teleported state is given by $C_{\rm out}(\lambda)=C_{\rm in}(\lambda)C_E(\lambda^*,\lambda)$ \cite{Marian}, and the fidelity is given by
\begin{eqnarray}
F=\frac{1}{\pi}\int d^2\lambda C_{\rm out}(\lambda)C_{\rm in}(-\lambda)
\end{eqnarray}
for a pure input state with $C_{\rm in}(\lambda)$. For the coherent-state inputs $|\gamma\rangle$, the fidelity does not depend on the initial amplitude $\gamma$, since the output amplitude always matches the input amplitude in the BK protocol. Thus, we have only to calculate the fidelity for one particular coherent state, e.g. vacuum state, to obtain the average fidelity for the whole set of coherent state inputs.
\begin{figure}
\centerline{\scalebox{0.25}{\includegraphics[angle=90]{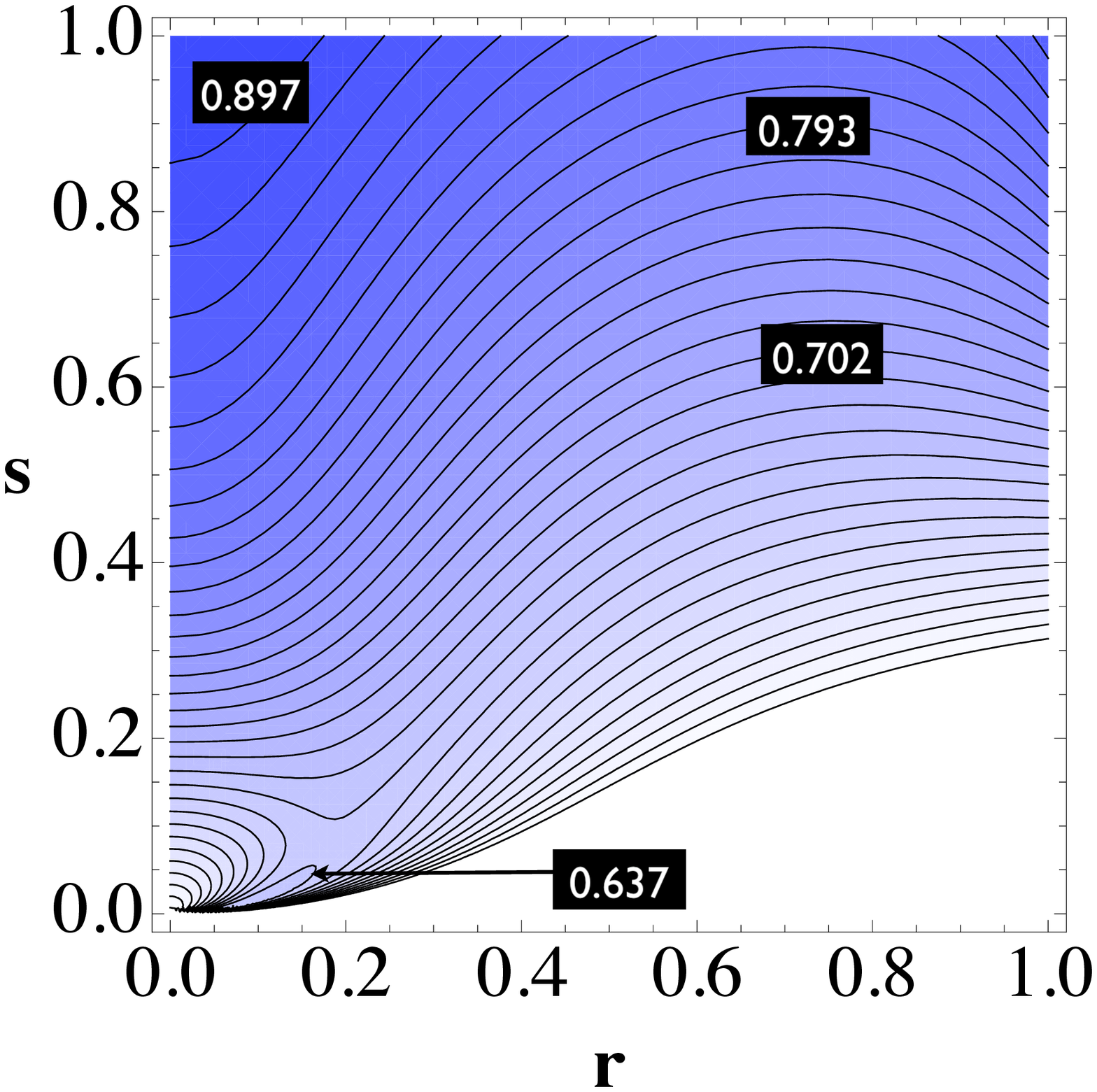}}}
\vspace{-0.2in}
\caption{(Color online) Average fidelity of teleporting a coherent state as a function of $s$ and $r$ for the state $(t\hat{a}+r\hat{a}^{\dag})(t\hat{b}+r\hat{b}^{\dag})|TMSS\rangle$. The colored region achieves the fidelity above the classical limit $1/2$.}
\label{fig:fig5}
\end{figure}
In Fig. 5, we show the fidelity of teleporting a coherent state using the resource via the coherent operation in a symmetric setting, $r_A=r_B=r$.  The maximum fidelity is achieved by the coherent superposition operation ($r\neq 0,1$) in the weak squeezing region, and the value of $r$ for the optimal fidelity approaches $r=0$ (photon subtraction) with the squeezing $s$, similar to the case of the EPR correlation.
The average fidelity generally increases with the squeezing parameter $s$. To overcome the classical limit $1/2$, $s$ must be larger than $s=0~(0.3047)$
at $r=0~(1)$. Interestingly, by comparing Figs. 2 and 5, we find that there exists a parameter region, e.g. $s=0.2$ and $r=0.5$, where no EPR correlation exists, nevertheless, the fidelity above the classical bound 1/2 is achieved. This suggests that the EPR correlation is not a necessary condition for a success of quantum teleportation beyond the Gaussian regime.

In Fig. 6 (a), we display the average fidelity in teleporting coherent states using various non-Gaussian resources. For the state $(t\hat{a}+r\hat{a}^{\dag})(t\hat{b}+r\hat{b}^{\dag})|TMSS\rangle$ via the coherent operation, we show the fidelity optimized over $r$ for each squeezing parameter $s$.
Compared with the other non-Gaussian states, the average fidelity is further improved by the coherent operation in the weak squeezing region.
For instance, we see from Fig. 6 (b) that the average fidelity$~0.65$ is achieved well above the classical bound 1/2 by the coherent operation,
for $s=0.01$, whereas it is a little over $0.5$ by the other non-Gaussian operations.
The non-Gaussian operation on one local mode only, $(t\hat{a}+r\hat{a}^{\dag})|TMSS\rangle$, which includes both the photon subtraction and the photon addition, does not improve the fidelity for any values of $r$.
For a moderate squeezing $0.075<s<0.417$, the operation $\hat{a}\hat{a}^{\dag}\hat{b}\hat{b}^{\dag}$ gives the best fidelity for a given squeezing $s$, whereas for a rather large squeezing $s>0.417$ the photon subtraction ($r=0$) becomes the optimal operation.

\begin{figure}
\centerline{\scalebox{0.35}{\includegraphics[angle=90]{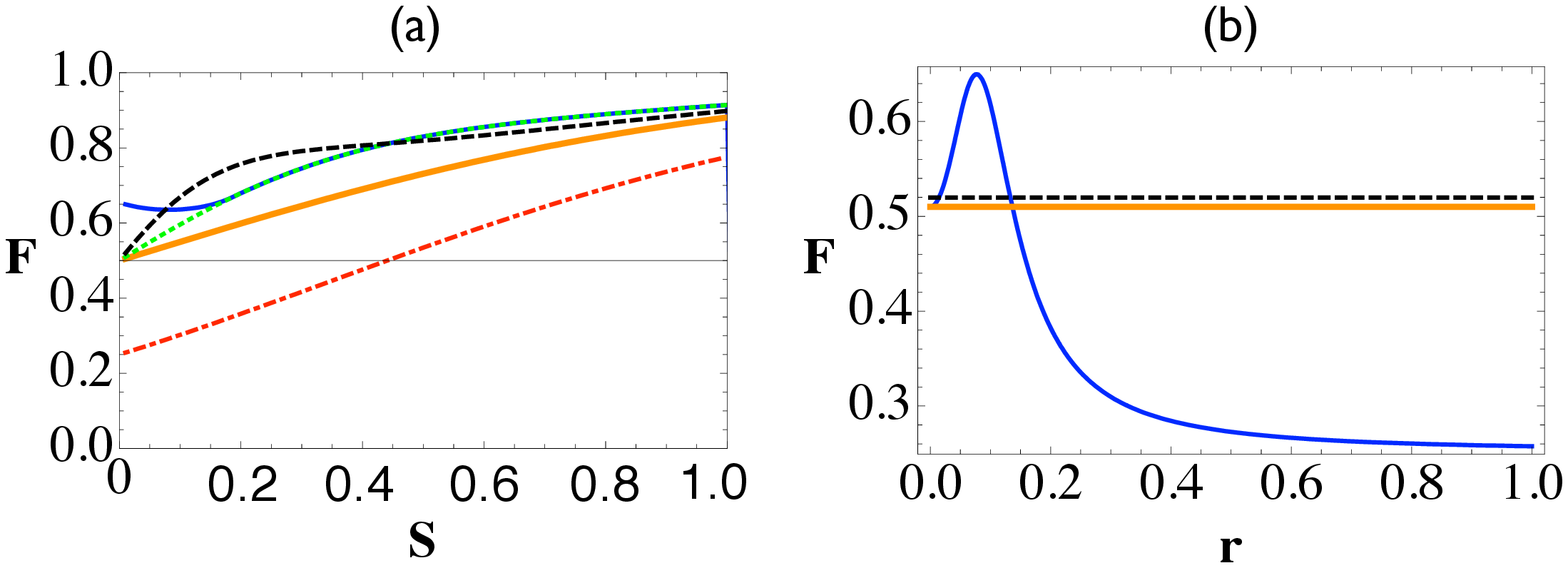}}}
\vspace{-1.5in}
\caption{(Color online) Average fidelity of teleporting a coherent state (a) as a function of $s$ and (b) as a function of $r$ with $s=0.01$ for the states: $|TMSS\rangle$ (orange thick solid), $\hat{a}|TMSS\rangle$ (red dot-dashed), $\hat{a}\hat{b}|TMSS\rangle$ (green dotted), $\hat{a}\hat{a}^{\dag}\hat{b}\hat{b}^{\dag}|TMSS\rangle$ (black dashed),  and $(t\hat{a}+r\hat{a}^{\dag})(t\hat{b}+r\hat{b}^{\dag})|TMSS\rangle$ (blue-solid curve). In (a), the value $r$ in the coherent operation is optimized for each $s$.}
\label{fig:fig6}
\end{figure}

\section{Conclusion}
In this paper, we have shown that the coherent superposition operation of photon subtraction and addition can enhance various entanglement characteristics of a two-mode state including the degree of entanglement, the EPR correlation, and the average fidelity of quantum teleportation. We have demonstrated that the coherent operation outperforms the previously studied non-Gaussian operations such as photon subtraction operation and photon-addition-then-subtraction operation in the weak squeezing regime, which may thus carry a practical significance.
In particular, it can be useful for the information processing using pulsed optical fields that typically show small squeezing. For instance, a recent experiment in Ref. \cite{Grangier} demonstrated 1 dB squeezing in the pulsed regime corresponding to the squeezing parameter $s\sim0.115$.  On the other hand, the optimal operation becomes the photon subtraction (case of $r=0$) in the large-squeezing regime.
The weak squeezing regime where the coherent operation gives a better result than the mere photon subtraction and the addition has been identified as (i) $s<0.44$ for the degree of entanglement, (ii) $s<0.135$ for the EPR correlation, and (iii) $s<0.17$ for the teleportation fidelity, respectively.

The present study can be further pursued to include other two-mode input states than the two-mode squeezed states and multiple coherent operations on each mode. In fact, for a fair comparison with other second-order operations, e.g. addition-then-subtraction $\hat{a}\hat{a}^\dag\hat{b}\hat{b}^\dag$,
the coherent operation can be applied twice, although we have here shown that the single coherent operations already beat $\hat{a}\hat{a}^\dag\hat{b}\hat{b}^\dag$ in the weak squeezing regime.
On another side, it seems necessary to work out a rigorous description of quantum teleportation in order to answer some unresolved questions, e.g., what is the sufficient or necessary condition to achieve a success of CV quantum teleportation beyond Gaussian regime.

This work is supported by the NPRP grant 08-043-1-011 from QNRF.
HN acknowledges research support by the Alexander von Humboldt Foundation.

\end{document}